\documentclass[12pt,preprint]{aastex}

\shorttitle{Kolmogorov--Burgers Turbulence}
\shortauthors{Boldyrev}

\begin{document}

\title{Kolmogorov--Burgers Model for Star Forming Turbulence}

\author{Stanislav Boldyrev\altaffilmark{1}}
\altaffiltext{1}{Email:  boldyrev@itp.ucsb.edu}
\affil{Institute for Theoretical Physics, Santa Barbara, CA 93106}

\begin{abstract}
The process of star formation in interstellar molecular clouds is 
believed to be controlled by driven supersonic 
magnetohydrodynamic turbulence. We suggest that in the inertial 
range such turbulence obeys the Kolmogorov law, while 
in the dissipative range it behaves 
as Burgers turbulence developing shock singularities. On the base 
of the She--L\'ev\^eque analytical model we then predict the velocity 
power spectrum in the inertial range to be~$E_k\sim k^{-1.74}$. This 
result reproduces the observational Larson  
law,~$\langle u^2_l \rangle\sim l^{0.74\cdots0.76}$, 
[Larson, MNRAS {\bf 194} (1981) 809] and agrees well  
with recent numerical findings by~Padoan and 
Nordlund [astro-ph/0011465]. 
The application of the model to more general 
dissipative structures, with higher fractal dimensionality, 
leads to  better agreement  with recent observational results.

\end{abstract}

\keywords{MHD: Turbulence --- ISM: dynamics --- stars: formation}
\section{Introduction}
\label{introduction}
It was recently argued on both observational and 
numerical grounds that star forming  regions of interstellar 
molecular clouds are governed by super-sonic and, possibly, 
super-Alfv\'enic  
turbulence, see, e.g.,~\citep{Padoan_Nordlund2}, 
(Padoan \& Nordlund 2000, 
hereafter~\citet{Padoan_Nordlund}),  
and a review  by~\citet{Elmegreen}.  
The turbulence is driven on large scales 
by supernovae explosions and energy is then transfered to  
smaller scales via a turbulent cascade, forming a 
hierarchy of dense clumps. It is still unclear whether such turbulent 
fragmentation is crucial on small scales, where Jeans-unstable 
density cores collapse and stars are formed. However, it seems reasonable 
that at least at the initial stage of a clumpy structure formation, 
turbulent fragmentation is the definitive process. This 
assertion, stemming from the work by~\citet{Larson},  was recently 
confirmed 
in the number of high-resolution numerical 
simulations~
\citep{Padoan_Nordlund,Padoan_etal,Mckee,klessen1,klessen2,Geyer_Burkert,Williams,MacLow2}. 

Observations suggest that the Mach number of turbulent motion,~$M$, 
can be greater than~10, and the Alfv\'enic Mach number,~$M_a$, 
can be greater than~1 [see, e.g.,~\cite{klessen1,Williams}]. 
Until recently, supersonic turbulence (both Navier-Stokes and MHD) 
has not received proper theoretical attention. 
In a series of 
papers, \citet{Porter_etal2} analyzed numerically 
decaying turbulence 
with initial Mach number of the order of~$1$. It has been observed in large 
resolution runs (up to $1024^3$) that the spectra of both the 
compressible and incompressible parts of the velocity 
field  approximately follow the Kolmogorov 
value,~$E_k\sim k^2 \vert u_k\vert^2 \sim k^{-5/3}$. 
However, decaying turbulence 
is different from forced turbulence in many aspects. To mention 
just a few, we
note that a supersonic motion forms shocks and quickly, on a 
crossing time,  
dissipates in decaying runs, while it can be sustained in forced ones. 
Also, it has been demonstrated by~\citet{Smith_etal1,Smith_etal2} 
that, in a decaying case, most energy 
is dissipated  in a large number of weak shocks 
contrary to a forced case where the largest shocks dissipate 
most of energy. In the present paper we consider supersonic, 
driven turbulent systems, stressing that they differ qualitatively 
from their subsonic, decaying counterparts.

In the last two years there appeared a number of papers analyzing 
numerically
forced supersonic turbulence both with and without magnetic fields. 
\citet{Porter_etal1} investigated forced non-magnetized turbulence 
with Mach number of the order of~$1$,  and observed no difference 
in  power spectra with the unforced runs. 
However, when~\citet{Padoan_Nordlund,Padoan_etal} simulated 
{\em supersonic} MHD turbulence ($M\sim 10$, $M_a\sim 3$), they 
found the velocity 
spectrum,~$k^{-\beta}$, with approximate value~$\beta=1.8$. This 
spectrum is steeper than the Kolmogorov one, which indicates strong 
intermittency effects. Correspondingly, velocity fluctuations scale 
with distance according to~$\langle u_l^2 \rangle \sim l^{\beta-1}$. 
The steeper-than-Kolmogorov spectrum was
linked to the supersonic nature of turbulence by~\citet{Larson1,Larson} 
on observational grounds. 

Our interest in supersonic turbulence is also motivated by the 
argument 
of~\citet{Padoan_Nordlund} that the spectral exponent,~$\beta$, 
may be directly related to the exponent of  
the mass distribution of collapsing cores,~$N(m)\sim m^{-1-\delta}$, 
as~$\delta = 3/(4-\beta)$. 
This  suggests that the initial mass function (IMF) could be
explained from the basic properties of turbulent fragmentation, 
without tunable parameters. The fact that supersonic 
MHD~turbulence leads to 
sustaining of shock turbulence, to shock fragmentation, 
and to establishing a certain universal density distribution 
has also been recently 
demonstrated by~\citet{Boldyrev_Brandenburg} in a one-dimensional 
solvable Burgers model. 


In this paper we present a theoretical model of driven  
supersonic turbulence, incorporating both Kolmogorov and Burgers 
pictures in different parts of the phase space.  We argue that 
due to mostly solenoidal character of such turbulence, 
the characteristic times of energy cascade in the inertial 
interval scale as in the Kolmogorov turbulence, 
while the dissipative structures are completely different. 
Instead of filaments, as in an incompressible
case, they can appear as sheets which is more 
consistent with Burgers turbulence. In Sec.~\ref{KB} we demonstrate 
that the standard She--L\'ev\^eque model~\citep{She_Leveque1,She_Leveque2},
which links the most singular turbulent structures with  turbulent 
spectra, has a solution corresponding to {\em sheet-like} 
dissipative structures, which reproduces the velocity power spectrum 
with exponent~$\beta=1.74$, close to the observational and 
numerical values. In Sec.~\ref{generalization} we generalize the 
results to more realistic, fractal dissipative structures, which 
leads to better agreement with recent observational results.

\section{Kolmogorov--Burgers model of supersonic turbulence}
\label{KB}
At first sight, turbulence with small pressure should behave in 
the same way as Burgers turbulence, the theory of which was 
substantially developed during the last few 
years~\citep{Polyakov,Yakhot,Boldyrev,Weinan,Gotoh,Eric,Verma,Frisch}. 
However, this is true only in one- and two-dimensional cases; in a 
three-dimensional case, the behavior of a compressible fluid
is qualitatively different from Burgers turbulence. The main 
difference is vorticity generation, an effect 
completely analogous to magnetic field generation existing in 
3D and non-existing in 2D. Indeed, the vorticity
equation,
\begin{eqnarray}
\partial_t {\bf \Omega} +({\bf u}\cdot \nabla) {\bf \Omega} - ({\bf \Omega} \cdot \nabla) {\bf u} + (\nabla \cdot {\bf u}) {\bf \Omega}
=\nu \Delta {\bf \Omega},
\end{eqnarray}
where the vorticity is~${\bf \Omega}=\nabla \times {\bf u}$, 
coincides with the induction equation for a magnetic field. 
Numerical experiments show that vorticity is generated quite 
effectively. In decaying turbulence with Mach numbers 
of the order of~$1$, simulated by~\citet{Porter_etal2}, 
it was found that the turbulence was mostly solenoidal.  
If one decomposes the velocity field into
the solenoidal part, $\nabla \cdot {\bf u_s}=0$, and the compressible part,
$\nabla \times {\bf u_c}=0$, their ratio was observed to 
be~$\gamma=\langle u_c^2 \rangle/\langle u_s^2\rangle \sim 0.1$ in the 
inertial range. A pressure term ensuring incompressibility in 
subsonic turbulence, turned out to be unimportant in  
supersonic dynamics: energy transfer over scales due to the pressure 
term was only~$3\%$. The subsequent {\em forced} runs 
by~\citet{Porter_etal1} revealed qualitatively the same results. 
In the case of forced turbulence with {\em large}  Mach 
numbers~$(M\sim 10, M_a\sim 3)$ and a solenoidal large-scale force, 
simulations by~\citet{Padoan_Nordlund} also
demonstrated that in the 
inertial interval 
this ratio is small,~$\gamma < 0.2$, but increases 
towards the dissipative region. This result is not 
sensitive to the character of the external force since compressible 
motion creates shocks and its divergent part decays faster than 
the solenoidal one [{\AA}ke Nordlund~(2001), 
private communication], however, it may be sensitive to 
the presence of magnetic field that is known to help generate
vorticity~\citep{Enrique}. 


These remarkable numerical observations lead us to a conjecture 
that the ratio~$\gamma$ can be treated as a small parameter in the 
theory of 3D~compressible turbulence. We assume that in the inertial 
region such turbulence is divergence-free, with the Kolmogorov time 
of velocity fluctuation,~$t_l\sim l^{3/2}$, where~$l$ is the size 
of the fluctuation.
Close to the dissipative range, shock structures start to play 
important role in energy transfer and dissipation. The turbulence in 
this region thus inherits certain properties of Burgers turbulence.
The theory allowing to link the most singular, dissipative structures of 
turbulence with its 
velocity spectrum was suggested by~\citet{She_Leveque1}. 
This theory represents a turbulent cascade as an infinitely 
divisible log-Poisson process that has three input parameters. Two of 
these parameters are naive scaling exponents,~$\Theta$ and~$\Delta$, 
of the velocity field and of the ``eddy-turnover time'', 
correspondingly: $u_l\sim l^{\Theta}$, $t_l\sim l^\Delta$. 
The other parameter is the co-dimension,~$C$, of the most singular 
dissipative structure (co-dimension is defined as dimension of space 
minus the dimension of the structure, $C=d-D$). The objective of the theory 
is to predict 
the so-called structure functions of the velocity field, defined~as 
\begin{eqnarray}
S_p(l) = \langle \left[ u(x+l)-u(x) \right]^p \rangle \sim l^{\zeta(p)}, 
\label{structure}
\end{eqnarray}
where~$u$ is a component of the velocity field parallel or 
transverse to~$l$. [According to the chosen component the structure 
functions are called either longitudinal or transversal. It is 
believed that both scale in the same way, so we do not 
specify what component is assumed in~(\ref{structure}).] 
The velocity spectrum is a 
Fourier transform of the second-order structure function and is 
given by~$E_k\sim k^{-1-\zeta(2)}$. If the turbulent cascade 
depended only on local eddy interactions, then the naive Kolmogorov scaling 
of structure functions would hold,  
$\zeta(p)=p/3$, and we would recover the energy 
distribution~$E_k\sim k^{-5/3}$. Real turbulence is however 
intermittent, which means that its spectrum is {\em not} 
determined by the naive scaling. The She--L\'ev\^eque theory  
predicts the scaling function~$\zeta(p)$ as
\begin{eqnarray}
\zeta(p)=\Theta \left(1-\Delta\right)p +C 
\left( 1- \Sigma^{p/3} \right), 
\label{structure_functions_general}
\end{eqnarray}
where~$\Sigma=1-\Delta/C$.
For the original derivation we refer the reader to 
the papers by~\citet{She_Leveque1,She_Leveque2,Dubrulle}; 
more practical discussion can be found 
in~\citep{Grauer,Politano,Biskamp}. 
For 3D~incompressible turbulence, the naive scaling exponents  
take the well-known Kolmogorov values~$\Theta=1/3$, and~$\Delta=2/3$, 
and the dissipative structures are known to be filaments, so 
their co-dimension is~$C=2$. With these input parameters, 
formula~(\ref{structure_functions_general}) reproduces experimental 
results for incompressible Navier-Stokes turbulence with 
accuracy of several percent up to~$p=10$. 

In our model of Kolmogorov--Burgers turbulence, 
the inertial range naive scaling exponents are Kolmogorov ones, while 
the dissipative structures are quasi-1D shocks, which gives~$C=1$ 
and~$\Sigma=1/3$.
Formula~(\ref{structure_functions_general})  now reads
\begin{eqnarray}
\zeta(p)=\frac{p}{9}+  1- \left(\frac{1}{3}\right)^{p/3}. 
\label{structure_functions}
\end{eqnarray}
This gives for the second-order structure 
function~$\langle u^2_l\rangle \sim l^{0.74}$, which reproduces 
the Larson law~\citep{Larson1,Larson}, and the velocity power spectrum is 
given by~$E_k\sim k^{-1.74}$, in a good agreement with
numerical results by~\citet{Padoan_Nordlund}. The intermittency 
correction to the Kolmogorov scaling is even larger for 
the {\em first-order} 
structure function,~$\langle |u_l| \rangle \sim l^{0.42}$, which can be  
checked observationally or numerically in an easier way. Our analysis 
here is analogous to the analysis of
{\em incompressible} MHD turbulence by~\citet{Grauer,Politano}, and 
also by~\citet{Biskamp}  who noted 
that the most singular
structures in such turbulence are micro-current sheets. The sheet-like 
dissipative structures together with the assumption that the energy 
cascade is given by the Kolmogorov rather than the Iroshnikov-Kraichnan 
mechanism, led the latter authors to the
same prediction for the structure function scaling as our
formula~(\ref{structure_functions}), which turned out to be in good 
agreement with numerical results. This indicates that both systems, 
though completely
different, belong to the same class of universality, in agreement 
with the ideas put
forward by~\citet{Dubrulle} and ~\citet{She_Leveque2}.

\section{Generalizations}
\label{generalization}

Our analysis relied considerably on {\em  sheet-like} shock structures. 
Analogous considerations for the
filament and core singularities would 
give~$E_k\sim k^{-1.697} $ for filaments~($C=2$), and~$E_k\sim k^{-1.685}$ 
 for cores~($C=3$). All these spectra are steeper than the Kolmogorov one. 
Although the shock-like dissipative structures are clearly seen in 
simulations, the Mach number and the resolution are not large enough to 
make precise comparison with molecular clouds.  As an important  
generalization of the theory, one can imagine that the dissipative 
structures are rather complicated on large scales and have 
dimensionality {\em greater} than two. This is consistent with observed 
fractal structures of the  {\em density} distribution, whose dimensionality 
is close to~$D=2.3$~\citep{Larson3,Elmegreen2,Chappell}. Since the 
substantial part of dissipation  occurs in shocks, see, 
e.g.,~\citep{Ostriker}, the 
dimension of the most singular dissipative structures may 
be close to~$D=2.3$ as well. 
[It would 
be interesting to check numerically the degree of correlation of 
the density field and the dissipation field.]
Substitution of~$C=0.7$ in our formula~(\ref{structure_functions_general}) 
leads to the first order structure function~$\langle |u_l| \rangle \sim l^{0.55}$, and to the energy 
spectrum~$E_k\sim k^{-1.83}$. The results of recent 
observations, see, e.g.~\citep{Brunt}, seem to agree well 
with these predictions.
However, more  precise measurement of the structure 
functions scaling would be required to 
indicate what structures are most important. The intermittency 
correction to a scaling exponent of the 
first-order structure function is large enough to 
be detectable in numerical experiments, but higher-order 
structure functions are more difficult to measure. An attempt to infer 
such structure functions from observations 
was made by~\citet{Miesch,Ossenkopf}, but the scaling was not 
established due to limited inertial ranges.


Another important question is the 
relation of the obtained spectrum to the initial 
mass distribution function. 
The consideration of~\citet{Padoan_Nordlund}  
was based on an 
implicit assumption of the mean-field 
approximation, while our explanation of the observed 
steeper-than-Kolmogorov spectrum is essentially based 
on {\em intermittency} effects. In the presence of strong fluctuations, 
this relation may be modified, also acquiring intermittency corrections. 

I am very grateful to  Richard Larson, {\AA}ke Nordlund, Annick Pouquet, 
Dmitri Uzdensky, and Enrique Vazquez-Semadeni for many 
valuable discussions, suggestions, and comments on both 
the physics and the style of 
the paper, and to Lars Bildsten, Bruce Elmegreen, 
Mordecai-Mark Mac Low, and John Scalo for useful remarks.


\end{document}